\documentclass[lettersize,journal]{IEEEtran}
\usepackage{amsmath,amsfonts}

\usepackage{array}
\usepackage{textcomp}
\usepackage{stfloats}
\usepackage{url}
\usepackage{verbatim}
\usepackage{graphicx}
\usepackage{cite}
\usepackage{booktabs}
\usepackage{svg}
\usepackage{multirow}
\usepackage{caption}
\usepackage{subcaption}
\usepackage[textsize=scriptsize]{todonotes}
\usepackage{cancel}
\usepackage{soul}
\usepackage{xcolor}
\usepackage{setspace}
\usepackage{tabularx}
\usepackage[normalem]{ulem}
\usepackage[ruled,vlined,linesnumbered,noend]{algorithm2e}
\usepackage{hyperref}

\usepackage{marginnote}

\newcommand{\etal}{\textit{et al.}}
\DeclareMathOperator*{\argmin}{arg\,min}

\begin{document}

\title{
Fast and interpretable electricity consumption scenario generation for individual consumers
}
\author{
        Jonas Soenen, 
        Aras Yurtman,
        Thijs Becker, 
        Koen Vanthournout, 
        Hendrik Blockeel 
    \thanks{
        Jonas Soenen, Aras Yurtman and Hendrik Blockeel are with KU Leuven, Department of Computer Science, Celestijnenlaan 200A box 2402, 3001 Leuven, Belgium; and Leuven.AI, Celestijnenlaan 200A box 2453, 3001 Leuven, Belgium. 
        Thijs Becker and Koen Vanthournout are with Flemish Institute for Technological Research (VITO), Boeretang 200, 2400 Mol, Belgium; and
        EnergyVille, Thor Park 8310, 3600 Genk, Belgium. 
    }
}

        % <-this % stops a space



\maketitle
\begin{abstract}
To enable the transition from fossil fuels towards renewable energy, the low-voltage grid needs to be reinforced at a faster pace and on a larger scale than was historically the case.  
To efficiently plan reinforcements, one needs to estimate the currents and voltages throughout the grid, which are unknown but can be calculated from the grid layout and the electricity consumption time series of each consumer. 
However, for many consumers, these time series are unknown and have to be estimated from the available consumer information. 
We refer to this task as scenario generation. 
The state-of-the-art approach that generates electricity consumption scenarios is complex, resulting in a computationally expensive procedure with only limited interpretability. 
To alleviate these drawbacks, we propose a fast and interpretable scenario generation technique based on predictive clustering trees (PCTs) that does not compromise accuracy. 
In our experiments on three datasets from different locations, we found that our proposed approach generates time series that are at least as accurate as the state-of-the-art while being at least 7 times faster in training and prediction.
Moreover, the interpretability of the PCT allows domain experts to gain insight into their data while simultaneously building trust in the predictions of the model. 
\end{abstract}

% \begin{IEEEkeywords}
% Article submission, IEEE, IEEEtran, journal, \LaTeX, paper, template, typesetting.
% \end{IEEEkeywords}

\section{Introduction}
\IEEEPARstart{T}{he} world is transitioning away from fossil fuels towards renewable energy sources to reduce greenhouse gas emissions. 
The European Commission has made binding declarations to achieve zero net greenhouse gas emissions by 2050 in the European green deal \cite{europeangreendeal}.
Because the production and use of energy accounts for more than 75\% of the EU's greenhouse gas emissions, decarbonizing the EU's energy system is critical to reach carbon neutrality \cite{europeangreendealenergy}. 
The United States recently approved the Inflation Reduction Act \cite{inflationreductionact}, a significant investment in the modernization of the American energy system. 
This legislation, combined with previous initiatives, puts the US on track to reduce their greenhouse gas emissions with an estimated 40\% by 2030 (compared to 2005)\cite{inflationreductionactemissionreduction}. 
Countries all around the world are setting similar targets\footnote{See \url{https://climateactiontracker.org/} for a comprehensive overview}. 

This large-scale transition will require significant investments on many fronts, one of which is the low-voltage grid (LVG).
Widespread adoption of electric vehicles, heat pumps, and photovoltaic (PV) panels will change electricity consumption and production habits significantly \cite{neaimeh_deakin_jenkinson_giles_2023}. 
The addition of these devices will increase the peak load on the LVG substantially, pushing it beyond its current limits. 
To handle future demand, the LVG must be reinforced \cite{ElementEnergy2022}. 
Moreover, it is anticipated that the pace of reinforcement will significantly surpass the present rate of renewal for the LVG \cite{Elmallah_2022}.

To plan such a reinforcement for the LVG effectively, one needs to know its current state, that is, the currents and voltages on individual cables and buses on fine-grained time scales (e.g., 15 min).  
These quantities are typically not measured for the entire grid due to its massive scale\footnotemark, and hence, need to be estimated with a model. 
\footnotetext{Installing the required measurement hardware on every feeder would be prohibitively expensive.}
Typical modeling approaches, such as a power flow calculation, require the electricity consumption time series of each connected consumer as input.

However, in Europe, for the vast majority of consumers, these electricity consumption time series are unavailable.
Not all consumers have smart meters installed \cite{Vitiello2022a}, the data might be measured but cannot be used due to privacy reasons, or it is too cumbersome to securely transmit, store and manage all the data on a central server.
Therefore, a distribution system operator (DSO) only has access to the electricity consumption time series of a limited set of so-called \emph{measured consumers} \cite{NavarroEspinosa2014}.
Additionally, DSOs typically have access to certain general (but often limited) attributes about every consumer such as yearly consumption and connection capacity.

One approach to address the data availability challenge is to use a model to predict the electricity consumption time series of an unmeasured consumer based on the available attributes of that consumer. 
This model is trained using the consumption time series data of the measured consumers alongside their respective attributes. 

There is an intrinsic stochasticity in the consumption under fixed circumstances; e.g., the time one cooks on a given day might randomly change although the available attributes remain the same. 
Hence, instead of making a single prediction given the available attributes, it is more desirable to have a model that captures this intrinsic stochasticity by generating multiple possible electricity consumption time series. 
We refer to this task as \emph{scenario generation}.

However, most of the existing scenario-generation techniques don't leverage the available consumer attributes to tailor the generated scenarios to specific consumers and circumstances. 
DSOs often use synthetic load profiles (SLPs) to model the electricity consumption of consumers for purposes such as billing. 
Such an SLP consists of a single time series for each consumer type which represents the \emph{average} electricity consumption of this type of consumer. 
Because typically only very few consumer types are defined, these SLPs are only very coarsely tailored towards individual consumers. 
Moreover, the SLPs themselves are not suited to analyze the LVG load and risk for congestion because modeling the average consumption vastly underestimates the peak load and doesn't capture the variability of the load.

Considering these limitations, existing LVG load studies avoid utilizing SLPs and instead rely on a collection of measured electricity consumption time series. 
To generate scenarios for a particular consumer, they simply sample randomly from the set of known time series~\cite{NavarroEspinosa2014, NavarroEspinosa2016, Veldman2015} (or from a probabilistic model learned from the known time series~\cite{Bernards2017, Bernards2020}). 
By generating multiple samples, these approaches capture the variability of the load and do not underestimate peak loads since every generated scenario is a real time series. 
However, the generation of these time series is not tailored towards specific consumers, which hinders predictive performance.

Nevertheless, generating scenarios that are tailored toward individual consumers and circumstances is critical to accurately estimate the state of the grid.   
For example, imagine 10\% of consumers have photovoltaic (PV) panels. 
If these consumers are concentrated in specific neighborhoods, they might generate a significant (local) load on the LVG, whereas these consumers will have less impact if they are spread evenly.
It is therefore important to take the consumer attributes into account to know which consumers have PV panels installed. 
Similarly, the circumstances also have a big influence; e.g., the electricity consumption time series of consumers with PV panels will be vastly different on a sunny day in summer compared to a rainy day in winter.

Soenen \etal{}~\cite{Soenen2023} and Azam \etal{}~\cite{Azam2023}, to our knowledge, the only models that leverage the available consumer, calendar, and weather attributes to generate scenarios tailored to particular consumers and circumstances. 
As is often the case, both methods generate scenarios by sampling from the time series of measured consumers. 
However, both of these methods are complex, resulting in computationally inefficient models with no or limited interpretability. There are inherent scalability challenges for these models, as we have found that it becomes impractical to employ them with extensive datasets of measured and/or unmeasured consumers.

In this paper, we propose a new approach to generate load scenarios based on any set of available attributes using a simple, fast, and interpretable model with similar predictive accuracy as the existing, more complex approaches. 
The two existing approaches consider clustering the electricity consumption time series and relating those clusters to the available attributes as two independent steps. 
Instead, in our proposed approach, we use predictive clustering trees (PCTs) that cluster the known time series and, \emph{at the same time}, relate these clusters to the available attributes.
In this way, the clustering is better aligned with the available attributes. 
Because our proposed method only consists of a single decision tree, it can easily be visualized, allowing the user to inspect which electricity consumption time series are predicted for which consumers and circumstances. 
This enables the user to gain insight into the driving factors behind the electricity consumption of individual consumers and builds trust in the predictions. 
Moreover, the proposed approach is drastically faster in both learning the model and generating scenarios; hence, it can be applied on large datasets. We have been able to apply this method to a (private) dataset of 170,000 profiles, a size that is large enough for a DSO to perform grid simulations with a representative set of profiles.

In the remainder of this paper, we first discuss the related work (Sec.~\ref{sec:related_work}), followed by the problem statement  (Sec.~\ref{sec:problem_statement}), and the explanation of the proposed method (Sec.~\ref{sec:methodology}). 
Afterwards, we describe our experimental methodology (Sec.~\ref{sec:experimental_methodology}) and present the experimental results (Sec.~\ref{sec:experiments}). 
Finally, we summarize the key takeaways in the conclusion (Sec.~\ref{sec:conclusion}).

\section{Related Work}
\label{sec:related_work} 
In this section, we first discuss several fields that are related but not directly applicable to electricity consumption scenario generation for unmeasured consumers. 
Afterward, we discuss the literature on scenario generation for unmeasured households in detail and compare it to our proposed methodology. 

\subsection{Related fields}
\label{sec:related_fields}
\paragraph{Forecasting} 
Time series forecasting techniques are not applicable to our setting because they predict future electricity consumption values given past electricity values of the same consumer~\cite{Shi2018, Reis2017, Vossen2018}. 
In the scenario generation setting that we consider, there are no historical consumption values available for the profiles that have to be predicted.

\paragraph{Residential load modeling}
A more closely related field is that of residential load modeling. 
The goal of residential load modeling is to model the process that generates the total electricity consumption of a residence or group of residences ~\cite{Proedrou2021, Swan2009, Grandjean2012}.
However, the existing techniques in this field are generally inapplicable for various reasons we discuss below. 
There are three categories of residential load modeling techniques: bottom-up, top-down, and hybrid techniques. 

Bottom-up techniques aggregate and extrapolate from fine-grained electricity consumption information, micro-variables (e.g., appliance-use data, occupant behavior, or individual consumer consumption profiles) to load patterns or profiles on a coarser level (e.g., a single consumer, a group of consumers, or a nation). 
Examples of this type of load modeling technique include Capasso \etal{}~\cite{Capasso1994} and Gottwalt \etal{}~\cite{Gottwalt2011}.
These techniques are generally inapplicable as the fine-grained electricity consumption information required for these techniques is not available to DSOs; this type of data is more difficult to gather than the electricity consumption profiles themselves. 

Top-down techniques use total electricity consumption time series and try to relate that electricity consumption to aggregate-level macro variables (e.g., household characteristics, weather, and macro-economic indicators) and/or stochastic predictors (arising from the analysis of historical time series data)~\cite{Proedrou2021}. 
Some of these top-down techniques can generate electricity consumption profiles for individual connection points. 
However, many of these techniques generate averaged or aggregated load profiles to represent the general trend of electricity consumption profiles~\cite{McLoughlin2015}. 
Similar to SLPs, these averaged or aggregated load profiles do not capture the full variability in load scenarios and are therefore unsuited as input to a power flow calculation. 
Other techniques do try to generate realistic load profiles but fail to capture the temporal correlation or can only leverage a single consumer attribute~\cite{toubeau2016} (or in some cases no attributes at all~\cite{Labeeuw2013}).

Hybrid techniques \cite{Johnson2014, Bartels1992} combine elements and methods from both bottom-up and top-down techniques. 
But these techniques are also inapplicable, as there are no micro variables available in our problem setting.

\paragraph{Generative neural networks}

There are a few papers that use generative neural networks architectures to conditionally generate load scenarios for individual residential customers. Generative adversarial networks (GANs) have been used to generate profiles conditioned on sociodemographic information \cite{LI2022124694,CHEN2023120711} or on clusters obtained from clustering the time-series \cite{8791575,WANG2020110299,CLAEYS2024122831}. A variational auto-encoder (VAE) has been used to generate scenarios conditioned on properties of the house (such as presence of an electric vehicle or smart tariff) \cite{faraday}. Generative neural networks have a few inherent disadvantages. 
First, training these networks is typically expected to be computationally intensive, which is of particular concern when models must be regularly retrained to accommodate dataset updates. 
Additionally, they often demand substantial volumes of data to achieve good performance, rendering them less applicable in scenarios with limited data availability (e.g., a DSO that only has a limited set of measured consumers). 
Moreover, these models provide minimal to no interpretability.  

\subsection{Existing work on scenario generation for unmeasured households} 

Soenen \etal~\cite{Soenen2023} and Azam \etal~\cite{Azam2023} proposed techniques that leverage all attributes (consumer, calendar, and weather) that are available to a DSO, to generate tailored scenarios. 
Both techniques sample scenarios from a dataset of historical consumption time series by using the available attributes to select the relevant historical time series. 
Here, we focus on the so-called ``data-driven'' technique proposed by Soenen \etal~\cite{Soenen2023}, as it is significantly faster and more scalable than the one proposed by Azam \etal~\cite{Azam2023}.
% }

The main idea behind the data-driven technique is to cluster the historical time series and to learn a probabilistic classifier that learns to predict which cluster a time series belongs to based on its associated attributes.
To generate a scenario for a given attribute vector, first, the classifier predicts which cluster these attributes belong to, and then, from that cluster, a random time series is sampled as the generated scenario. 
The drawback of this two-step approach is that clustering and the subsequent classification are performed independently.
Therefore, the clustering might not be aligned with the available attributes, making it difficult for the probabilistic classifier to assign attribute vectors to clusters. 
This simple approach would entail clustering all historical daylong time series, which is computationally expensive. 
Hence, the data-driven approach applies the clustering and classification procedure twice in a two level-approach.
In the first level, it uses the household attributes to select a cluster of yearlong time series and then, in the second level, within the selected cluster of yearlong time series, the daily attributes are used to select a cluster of daylong time series to sample from.  

Our proposed approach shares a similarity with the data-driven approach of Soenen \etal{}~\cite{Soenen2023} in that both approaches sample scenarios from a set of historical time series. However, we perform the sampling differently by using a PCT. A PCT integrates clustering and classification by grouping historical time series into clusters that are distinguishable by their attributes. This ensures that the clustering is aligned with the attributes, unlike the data-driven approach that performs clustering and classification separately. Additionally, the PCT-based approach is efficient enough to operate on a single level, unlike the data-driven approach that needs a two-level procedure to be computationally viable, making it less optimal. Unlike the data-driven approach, our proposed approach is interpretable, as discussed in Section \ref{sec:exp_interpretability}.

\section{Problem Statement} 
\label{sec:problem_statement}
\noindent

\textbf{Given:} A dataset $\mathcal{D}$ that contains $N$ historical daylong electricity consumption time series $\mathbf{t}_i$ each associated with an attribute vector $\mathbf{a}_i$: 
\begin{equation} \label{eq:dataset}
    \mathcal{D} = \{(\mathbf{a}_i, \mathbf{t}_i)~:~i=1, \dots, N \}.
\end{equation}
The attribute vector $\mathbf{a}_i$ contains all attributes that are available for the scenario generation task such as consumer attributes (e.g., yearly consumption), weather attributes (e.g., temperature), and calendar info (e.g., day of week). \\

%\noindent 
\textbf{Do:} Generate $N_S$ load scenarios $\{\mathbf{\hat{t}}^{(s)}~:~s=1, \dots, N_s \}$, i.e., daylong electricity consumption time series, for a specific attribute vector $\mathbf{a}'$, where $N_s$ is the number of scenarios to be generated which is specified in advance. 

\section{Methodology}
\label{sec:methodology}
Our methodology uses a predictive clustering tree (PCT), a machine-learning model that combines elements from both clustering and prediction \cite{Blockeel1998, Dzeroski2007}.
A PCT is a decision tree where splits are based on the attributes in $\mathbf{a}_i$ and each leaf node represents a cluster of time series $\mathbf{t}_i$.
A PCT can be learned with standard greedy top-down decision tree induction where splits on $\mathbf{a}_i$ are chosen to minimize the intra-cluster variance of the time series $\mathbf{t}_i$ in each child node (Sec.~\ref{sec:method_training}). 
Minimizing the intra-cluster variance leads to a clustering where time series with similar consumption patterns end up in the same leaf (or cluster) and dissimilar time series end up in different clusters. 
The learned decision tree allows us to predict the cluster of an arbitrary attribute vector $\mathbf{a}'$. 
Because clustering and building the prediction model happen simultaneously, the clusters are constructed in such a way that they are separable in the space of all attributes and are, therefore, well-suited for prediction.

To generate scenarios for a given attribute vector $\mathbf{a}'$, the attribute vector $\mathbf{a}'$ is assigned to one of the clusters by traversing the tree from root to leaf following the edges that are satisfied for $\mathbf{a}'$; the scenarios are then sampled randomly from the selected cluster (Sec.~\ref{sec:method_generation}). 

\subsection{Building a predictive clustering tree}
\label{sec:method_training}
Building a predictive clustering tree is similar to standard greedy top-down decision tree induction used in a number of algorithms such as C4.5 \cite{quinlan2014}.
Starting with the root node that contains all instances, nodes are recursively split until no suitable split can be found or some stopping criterion is reached (e.g., maximum tree depth). 
Pseudocode for this procedure can be found in Algorithm~\ref{alg:PCT} where lines 1--9 show the recursive splitting procedure and line 16 the stopping criterion. 

To find the best way to split a node $D$, the tree induction algorithm (lines 11-18 in Algorithm~\ref{alg:PCT}) exhaustively searches for the splitting condition on an attribute in $\mathbf{a}$ to minimize the variance of the time series within each resulting child node. 
More specifically, the tree induction algorithm chooses the split that results in the highest variance reduction. 
The variance reduction $h$ for splitting a node $D_P$ into the partition $\mathcal{P}$ that contains child nodes $\{D_j\}$ is 
\begin{equation}
    h(D_P, \mathcal{P}) = \mathrm{var}(D_P) - \sum_{D_j \in \mathcal{P}}\frac{|D_j|}{|D_P|} \mathrm{var}(D_j)
    \label{eq:variance_reduction}
\end{equation}
where $\mathrm{var}(\cdot)$ denotes the variance of a node and $|\cdot|$ the number of instances in a node. 
A negative variance reduction corresponds to a split that increases the variance, and hence, is not considered as a valid split.

Given an arbitrary distance metric $\mathrm{dist}(\cdot)$ between time series, the variance of a node $D$ can be computed in two ways~\cite{Dzeroski2007}: 
\begin{itemize} 
 \item in $O(N^2)$ distance computations, using the sum of squared pairwise distances:%\todoaras{equation corrected}
\begin{equation}
    \mathrm{var}(D) = \frac{1}{2|D|^2} \sum_{(\mathbf{a}_i, \mathbf{t}_i) \in D} \; \sum_{(\mathbf{a}_j, \mathbf{t}_j) \in D} \mathrm{dist}(\mathbf{t}_i, \mathbf{t}_j)^2
    \label{eq:variance_pairwise}
\end{equation}
 \item or, in $O(N)$ distance computations, if the centroid of a node can be efficiently computed, as the average squared distance to the centroid $\overline{D}$:
\begin{multline}
    \mathrm{var}(D) = \frac{1}{|D|} \sum_{(\mathbf{a}_i, \mathbf{t}_i) \in D} \mathrm{dist}(\mathbf{t}_i,\overline{D})^2 \\
    \text{with } \overline{D} = \argmin_{\mathbf{x}} \sum_{(\mathbf{a}_i, \mathbf{t}_i) \in D } \mathrm{dist}(\mathbf{x}, \mathbf{t}_i)^2.
    \label{eq:variance_centroid}
\end{multline}
\end{itemize}

In this paper, we use the Euclidean distance for scalability, as most datasets in this field contain $\approx 10^6$ time series (Sec.~\ref{sec:datasets}).
Under the Euclidean distance, the centroid $\overline{D}$ of a node can be computed in linear time using: 
\begin{equation}
     \overline{D} = \frac{1}{|D|} \sum_{(\mathbf{a}_i, \mathbf{t}_i) \in D } \mathbf{t}_i.
\end{equation}
Therefore, using Eq.~\ref{eq:variance_centroid}, the variance can be computed in linear time as well. 
On the other hand, for many time-series-specific distance metrics, no closed-form expression for the centroid $\overline{D}$ is known where one needs to resort to (iterative) approximate algorithms or an exact calculation (equation \ref{eq:variance_pairwise}) that has quadratic time complexity in the number of instances. 
For example, for the dynamic time warping distance\cite{Sakoe1978} between time series the centroid needs to be approximated using dynamic time warping barycenter averaging, which is not guaranteed to converge to the global optimum\cite{Petitjean2011a}. 

We only consider numerical attributes, therefore, all splits are binary splits of the form  $attribute < threshold$ where the attribute is one of the attributes in $\mathbf{a}$ and $threshold \in \mathbb{R}$. 

To avoid overfitting, we first learn a tree using the majority of the available data $\mathcal{D}$, constraining the tree by setting the maximum depth and the minimum samples per leaf.
Afterwards, we prune this tree using the remainder of $\mathcal{D}$ (the so-called pruning set) with reduced error pruning\cite{Quinlan1987}, 
which removes splits from the tree that do not reduce the error, in our case, variance, on the pruning set. 
This pruning step reduces the importance of the maximum depth hyperparameter, provided that it is set to a sufficiently large value such that pruning limits the final depth of the tree.

\begin{algorithm}[b]
\caption{PCT induction algorithm}\label{alg:PCT}
\small
% Don't print semicolons
\DontPrintSemicolon
% define function name
\SetKwFunction{FTree}{build\_tree}
\SetKwFunction{FBestSplit}{best\_split}
\SetKwFunction{FAdmissible}{allowed}
\SetKwProg{Pn}{function}{}{}
\Pn{\FTree{$D$}}{
        $split, \mathcal{P}^*  \gets$\FBestSplit{$D$}\;
        \eIf{$split \neq None$}{
            subtrees $\gets$ [] \;
            \For{$D_i \in \mathcal{P}$}{
                $subtrees[i] \gets $\FTree{$D_i$}\;
            } 
            \KwRet{node($split$, subtrees)}\;
        }{
            \KwRet{leaf($D$)}\;
        }
}
\;

\Pn{\FBestSplit{$D$}}{
    $h^*, s^*, \mathcal{P}^* \gets 0, None, None$ \;
    \For{each possible split $s$}{
        $\mathcal{P} \gets $ partition induced by split $s$ on $D$\;
        $h \gets \mathrm{var}(D) - \sum_{D_i \in \mathcal{P}} \frac{|D_i|}{|D|} \mathrm{var}(D_i)$ \;
        \tcc*{\footnotesize Allowed checks whether the split is allowed according to the tree building constraints (e.g., maximum depth)}
        \If{$h>h^* \wedge$ \FAdmissible{test, $\mathcal{P}$}}{
            $h^*, s^*, \mathcal{P}^* \gets h, s, \mathcal{P}$ \;
        }
    }
    \KwRet{$s^*, \mathcal{P}^*$}
}

\end{algorithm}

\subsection{Scenario generation} 
\label{sec:method_generation}
After learning a PCT, generating scenarios for a certain attribute vector $\mathbf{a}'$ is straightforward:
First, assign the attribute vector to one of the leaves in the tree by traversing the tree and, at each split, selecting the child node whose condition is satisfied for $\mathbf{a}'$. 
Second, as this leaf corresponds to a cluster of time series $\{\mathbf{t}_i\}$, randomly sample $N_s$ scenarios and return these as generated scenarios $\mathbf{\hat{t}}^{(s)}$ for $s = 1, \cdots, N_s $. 
The simplicity of this procedure makes the model interpretable, which will be explored in the experimental results section (Sec.~\ref{sec:exp_interpretability}).

\section{Experimental methodology}
\label{sec:experimental_methodology}
In this section, we describe the details of our experiments: the three datasets we used (Sec.~\ref{sec:datasets}), the cross-validation scheme (Sec.~\ref{sec:crossval}), the implemented scenario generation techniques (Sec.~\ref{sec:methods}), and the performance metric (Sec.~\ref{sec:performance_metric}). 

\subsection{Datasets} 
\label{sec:datasets}
We use three datasets of load time series acquired from three different locations: Flanders (a region in Belgium), Ireland, and London. 
The datasets from Ireland and London are publicly available (upon request)\cite{irishdata, londondata}.
Every dataset is enriched with publicly available weather data and calendar information (Sec.~\ref{sec:data_enrichment}). 
We consider the Flanders dataset as the most realistic dataset for our problem setting, because the available attributes are in fact known by the DSO, whereas the Ireland and London datasets contain more detailed attributes that are difficult to collect on a large scale.
However, these publicly available datasets do facilitate comparison with other methods and make our results reproducible.

We learn and evaluate the models directly on the unnormalized data as we are not only interested in the shape of the generated scenarios but also their scale. 
The data-driven and expert-based approach proposed by Soenen \etal{}~\cite{Soenen2023} require full yearlong time series for every consumer. 
To ensure compatibility and a fair comparison, every dataset is preprocessed to retain only full yearlong time series.
To ensure reproducibility, the code of the proposed method, the dataset details and all dataset preprocessing code for the two open datasets are available online\footnote{\url{https://github.com/jonassoenen/predclus_scengen}}. 
An overview of the dataset characteristics, after preprocessing, is given in Table~\ref{table:dataset_characteristics}.
In what follows, we describe each dataset in detail.

\begin{table*}
\center\footnotesize
\caption{Characteristics of the datasets used in this paper (detailed descriptions are provided in Sec.~\ref{sec:datasets})}
\label{table:dataset_characteristics}
\begin{tabular}{lllllll}
\toprule
Dataset & \# yearlong time series & \# consumers & sampling period & missingness & measurement years & availability\\
\midrule
Flanders & 4363                    & 2200                & 15 min  &0.30\%      & 2010-2017  & private       \\
Ireland   & 3488                    & 3488                & 30 min  &  0.05\%     & 2010      & upon request\cite{irishdata}       \\
London  & 2621                    & 2476                & 30 min    &   0.28\% & 2012-2013    & upon request\cite{londondata}   \\
 \bottomrule
\end{tabular}

\end{table*}

\subsubsection{Flanders dataset} 
\label{sec:data_flanders}
The dataset from Flanders is acquired from residential consumers and businesses connected to the LVG by Fluvius, the DSO in Flanders. 
The dataset consists of 4363 yearlong time series sampled every 15 min that belong to 2200 unique consumers between 2010 and 2017. 
The available attributes are yearly consumption (in kWh); connection capacity (in kVA), i.e., the maximum power that a consumer can draw from the grid; whether or not PV panels are installed; and PV capacity (in kVA), i.e., the maximum power that the installed PV panels can produce. 
Because of privacy reasons, this dataset cannot be made publicly available.

\subsubsection{Ireland dataset} 
\label{sec:data_ireland}
The publicly available Ireland dataset originates from the Smart Metering Electricity Consumer Behaviour Trials executed by the Commission for Energy Regulation (CER)\cite{irishdata}. 
The dataset contains 30-min load measurements from over 5000 Irish homes and businesses from July 2009 until December 2010 with detailed survey data for every consumer. 
Because the survey data of homes and businesses are different and cannot be consolidated into a single attribute set, we focused on the 3488 residential consumers in the dataset.
The survey data is filtered and preprocessed into 33 attributes relating to the inhabitants, building, home heating type, water heating type, cooking energy source, and yearly electricity consumption.

\subsubsection{London dataset} 
\label{sec:data_london}
The publicly available London dataset originates from the Dynamic Time-of-Use Electricity Pricing trial part of the Low Carbon London Project \cite{londondata}.
The dataset contains 30-minute load measurements from November 2011 until February 2014 complemented with survey data of every consumer. 
Discarding incompletely measured years results in 2621 yearlong time series from 2476 unique consumers where all consumers have data in 2013 and a few consumers have data for both 2012 and 2013.
The survey data is filtered and preprocessed into 75 attributes relating to the inhabitants, building, insulation, home heating type, water heating type, appliances, and yearly electricity consumption.

\subsubsection{Data enrichment with weather and calendar attributes} 
\label{sec:data_enrichment}
Every dataset is enriched with 8 weather attributes and 7 calendar attributes. 
The weather attributes are: minimum, maximum and average temperature during the day; average feels-like temperature during the day (all temperatures are in $^\circ \text{C}$); sun hour, i.e., a measure for the amount of solar radiation; the UV index; concentration of water vapor in the air (in \%); and wind speed (in km/h).
The historical weather attributes were obtained from WorldWeatherOnline\footnote{\url{https://www.worldweatheronline.com/}}.
Because the location of the individual consumers is unknown for the Flanders and Ireland datasets, we used weather information from Brussels and Dublin, respectively.
For the London dataset, we simply used weather information from London. 

The calendar attributes are day of week, day of month, day of year, month, season, whether or not it is weekend, and whether or not it is a holiday. 
These attributes are derived from the date itself except for the holiday information which is obtained with  the ``holidays'' Python package\footnote{\url{https://pypi.org/project/holidays/}}.

\subsection{Cross validation} 
\label{sec:crossval}

To measure the performance of each scenario generation technique, we apply 5-fold cross-validation. 
For each dataset, the unique consumers are randomly partitioned into 5 approximately equal-sized folds\footnote{This ensures that consumers whose load is being predicted are not included in the training set (avoiding leakage \cite{leakage}).}. 
The data from four of these folds are used as a training set while the remaining fold is used as a test set. 
Models are trained based on the tuples $\{(\mathbf{a}_i, \mathbf{t}_i)\}$ in the training set. 
For every tuple $(\mathbf{a'}, \mathbf{t})$ in the test set, the trained model generated $N_s$ scenarios $\{ \mathbf{\hat{t}}^{(s)} \}$ based on $\mathbf{a'}$. 
These scenarios are then compared with the ground truth $\mathbf{t}$ using the energy score (explained in Sec.~\ref{sec:performance_metric}).  
This process is repeated five times such that each fold is used for testing exactly once.

\subsection{Performance metric}
\label{sec:performance_metric}
Evaluating the quality of the generated scenarios is not straightforward. 
If exactly the same circumstances $\mathbf{a}'$ would occur multiple times, we still expect the electricity consumption time series of a specific consumer to vary considerably due to the inherent stochasticity of human behavior and unobserved variables. 
In our problem setting, it is important that the generated scenarios based on $\mathbf{a}'$ reflect this variance. 
If a standard performance metric such as mean squared error were used to measure the quality of the generated scenarios, consistently predicting the expected value would be optimal; but this would underestimate the possible variation, which is undesired in our problem setting. 
Furthermore, since electricity consumption time series typically have many peaks, evaluation using the mean squared error suffers from the double peak penalty effect \cite{Haben2014}.
Therefore, the mean squared error is not a suitable evaluation metric.

The Energy Score (ES)~\cite{Gneiting2008} is a suitable measure to assess the quality of the generated scenarios \cite{Moehrlen2023}. 
The ES~\cite{Gneiting2008} is a multivariate generalization of the conditional ranked probability score (CRPS) \cite{Matheson1976} which assesses the quality of a multivariate probabilistic forecast. 
The ES is a strictly proper score~\cite{Broecker2007} that assesses both the calibration and sharpness of the predicted multivariate distribution\cite{Gneiting2008}. 
In the case of scenario generation, where a set $\left\{\hat{\mathbf{t}}^{(s)}\right\}_{s = 1, \dots, N_s}$ of equally likely scenarios is predicted for a given test sample $\mathbf{t}$, the ES can be calculated as 
\begin{multline} 
    \mathrm{ES}\!\left(\left\{{\mathbf{t}'}^{(s)}\right\}_{s = 1, \dots, N_s},\; \mathbf{t}\right) \\ =
    \frac{1}{N_s} \sum_{s = 1}^{N_s} \left\Vert {\mathbf{t}'}^{(s)} - 
    \mathbf{t} \right\Vert - \frac{1}{2 N_s^2} \sum_{s = 1}^{N_s} \sum_{r=1} ^{N_s} \left\Vert {\mathbf{t}'}^{(s)} - {\mathbf{t}'}^{(r)} \right\Vert
\end{multline}
where $||\cdot||$ denotes the Euclidean norm. 
We report the average ES over all the daylong time series in the test set, where a lower ES means more accurate scenarios.

\subsection{Scenario Generation Methods} 
\label{sec:methods}
In our experiments, we evaluate four scenario generation techniques: 

\begin{itemize} 
\item \emph{Random sampling (baseline)}: Selecting scenarios randomly from the daylong time series in the training set, without taking into account attributes, as observed in existing LVG studies~\cite{NavarroEspinosa2014, NavarroEspinosa2016, Veldman2015}.
\item \emph{The expert-based technique} proposed by Soenen \etal{}~{\cite{Soenen2023}}.
We use the same configuration as \cite{Soenen2023}: the consumers are clustered based on the normalized yearly consumption and connection capacity. The number of clusters is automatically selected from \{5, 10, \dots, 100\} by the elbow method\cite{Satopaa2011}. 
\item \emph{The data-driven technique} proposed by Soenen et al.~\cite{Soenen2023}. 
Again, we use the same configuration as the authors:  the number of clusters is automatically selected from \{5, 10, \dots, 100\} with the elbow method\cite{Satopaa2011} using the Random Forest Classifier\cite{Breiman2001} from scikit-learn\cite{Pedregosa2011} with default hyperparameters. 
\item \emph{The proposed predictive clustering approach} described in Sec.~\ref{sec:methodology}.
75\% of the training set is used to build the PCT with a maximum depth of 12 and minimum 300 samples per leaf. 
The remaining 25\% is used as a pruning set (to prune the PCT). 
We learn PCTs using the Clus software package\footnote{\url{https://dtai.cs.kuleuven.be/clus/}}. 

\end{itemize}

For each daylong time series in the test set, each technique generates 250 daylong load scenarios ($N_s = 250$). 
The main experiments were run on a linux server with 128 GB of RAM and 2 Intel(R) Xeon(R) Silver 4214 CPUs with 12 cores and 24 threads each. 
Part of the data-driven and expert-based method are multi-threaded, giving them a significant advantage over the proposed approach when measuring runtimes.

\section{Experimental results} 
\label{sec:experiments}

\subsection{Scenario generation performance}\label{sec:scengenperf}
We compare the scenario generation performance of the proposed predictive clustering approach to the expert-based and data-driven approaches\cite{Soenen2023}, and to a baseline of random sampling. 
The comparison is performed based on the Flanders, London, and Irish datasets using the average ES over all cross-validation folds as the performance metric. 
Because the expert-based approach requires the connection capacity as an attribute, it can only be executed on the Flanders dataset. 
To compare the execution time of the methods, we measured the training time (the time it took to train the model on the training set) and the scenario generation time (the total time it took to generate the scenarios for the test set). 

\begin{figure}[t]
    \centering
    \includegraphics[width = 0.78\linewidth, trim={0 0 0 1.5cm}]{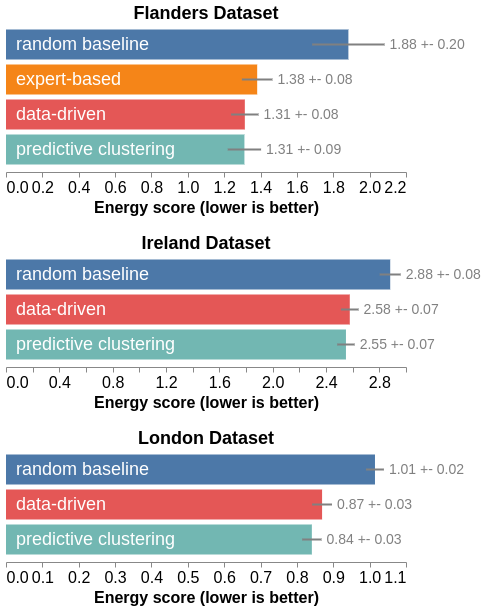}
    \caption{The proposed predictive clustering technique performs on par with the existing data-driven approach, and, on the London dataset, slightly outperforms it. The bars show the average ES over all folds, the gray lines shows the standard deviation of ES over the folds.}
    \label{fig:general_comparison}
\end{figure}

\textit{Results} Our proposed predictive clustering approach generates scenarios that are as accurate as the state-of-the-art data-driven technique, and, on the London dataset, even slightly more accurate than the data-driven technique (Fig.~\ref{fig:general_comparison}). 
In terms of training time, the predictive clustering approach is on average 10 times faster than the data-driven method, but significantly slower than the expert-based method (Table~\ref{tab:overal_comparison_runtimes}).
In terms of scenario generation time, predictive clustering is by far the fastest method. 
Prediction is very slow for the data-driven method (taking up to 6 hours for the Ireland dataset) as it needs to make multiple predictions with a random forest.

\textit{Conclusion} 
In terms of energy score, our (simple) predictive clustering approach performs on par with, or slightly better than, the complex data-driven approach while being significantly faster in both training and scenario generation time. 

\begin{table}[t]
\caption{For every dataset, the average training time and prediction (i.e., scenario generation) time over every fold.}
\label{tab:overal_comparison_runtimes}
\centering\small
\begin{tabular}{llrr}
\toprule
 & & \multicolumn{2}{c}{\textbf{Runtime (in sec)}} \\
  % \cmidrule{3-4}
  \textbf{Dataset} & \textbf{Method}   &  \textbf{Training} & \textbf{Prediction} \\
  \midrule
\multirow[t]{3}{*}{Flanders} & data-driven & 6388  & 7505  \\
 & expert-based & 70 & 4087 \\
 & predictive clustering & 454 & 96 \\
 \addlinespace[0.5em]
\multirow[t]{2}{*}{Ireland} & data-driven & 3707 & 24372 \\
 & predictive clustering & 340 & 77 \\
 \addlinespace[0.5em]
\multirow[t]{2}{*}{London} & data-driven & 2966 & 19718 \\
 & predictive clustering & 407 & 58 \\
 \bottomrule
\end{tabular}

\end{table}

\subsection{Influence of dataset size} 
\label{sec:exp_influence_dataset_size}

\begin{figure*}
    \hfill%
    \begin{subfigure}[b]{0.48\textwidth}
        \includegraphics[width = \linewidth, trim={0 0 0 1.5cm}]{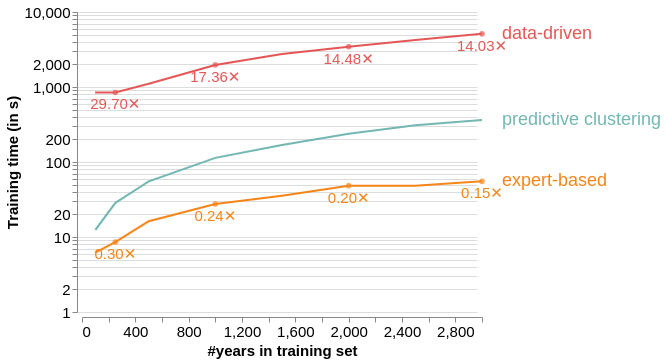}
        \caption{Time to train each model }
        \label{fig:dataset_influence_training}
    \end{subfigure}
    \hfill%
    \begin{subfigure}[b]{0.48\textwidth}
        \includegraphics[width = \linewidth, trim={0 0 0 1.5cm}]{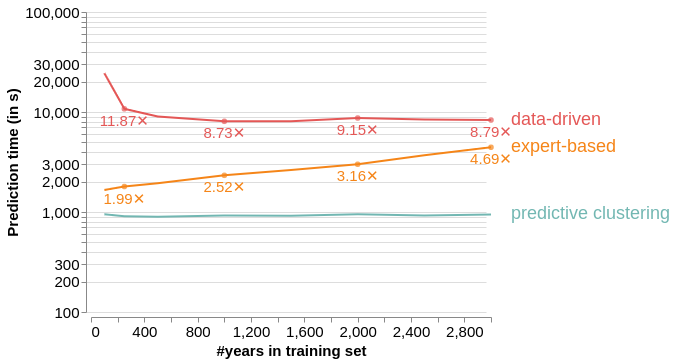}
        \caption{Time to generate scenarios for test set}
        \label{fig:dataset_influence_testing}
    \end{subfigure}
    \hfill%
    \caption{Training and prediction times as a function of training set size for every method on the Flanders dataset. For the data-driven and expert-based method, the relative execution time compared to predictive clustering is shown as well (e.g. for 250 years in the training set, data-driven needs 29.7 times more time to train a model than the proposed approach).  }
    \label{fig:dataset_influence_time}
\end{figure*}

In this experiment, we study the influence of the dataset size on the performance and the execution time of every method. 
To achieve this, a test set of $500$ yearlong time series is randomly sampled from the Flanders dataset.  
Next, multiple training sets of sizes [100, 250, 500, 1000, 1500, 2000, 2500, 3000] are sampled from the full dataset, ensuring each smaller dataset is a subset of the next (bigger) dataset.
Then, every method is trained with every training set and evaluated on the fixed test set with the energy score. 
As before, we also measure the training and scenario generation time.\footnote{To manage computation time, the calculations for this experiment were performed on different servers than the main experiments that use full datasets; hence, the training and prediction times shown in Fig.~\ref{fig:dataset_influence_time} should not be directly compared to those in Table~\ref{tab:overal_comparison_runtimes}. The key takeaway from Fig.~\ref{fig:dataset_influence_time} is the relative differences between the runtimes.}

\begin{figure}[t]
    \centering
    \includegraphics[width = .85\linewidth, trim={0 0 0 1.5cm}]{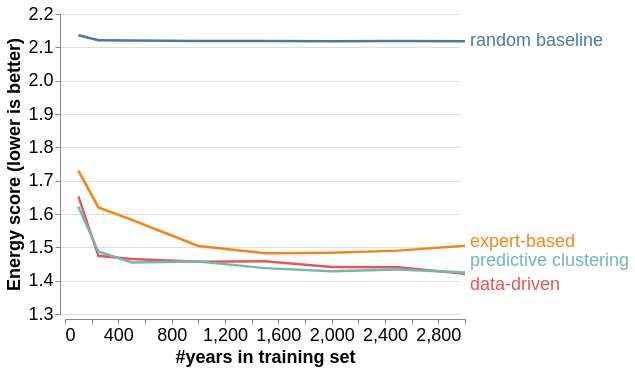}
    \caption{Evolution of the energy score as a function of training set size for every scenario generation method on the Flanders dataset.  }
    \label{fig:dataset_influence_ES}
\end{figure}
\textit{Results}
In terms of ES, both predictive clustering and data-driven techniques keep producing better and better scenarios as the number of yearly profiles in the training set grows (Fig.~\ref{fig:dataset_influence_ES}). 
However, most of the ES reduction happens between 100 and 500 yearlong time series in the training set. 
The expert-based method keeps improving up to 1500 yearlong time series but never reaches the same ES as the other two techniques. 
The random baseline does not benefit from a bigger training set; its ES is almost constant around $2.11$.
In terms of training time, predictive clustering is more than 10 times faster than the data-driven method but slower than the expert-based method (Fig.~\ref{fig:dataset_influence_training}). 
To generate predictions for $500 \times 365$ days, predictive clustering is the fastest method regardless of the training set size (Fig.~\ref{fig:dataset_influence_testing}).

\textit{Conclusion}
Both predictive clustering and the data-driven method benefit from larger training sets, albeit with diminishing returns.

\subsection{Interpretability}
\label{sec:exp_interpretability}
\begin{figure*}
    \centering
    \includegraphics[height =0.75\textheight, trim={0 2.5cm 0 4.5cm}]{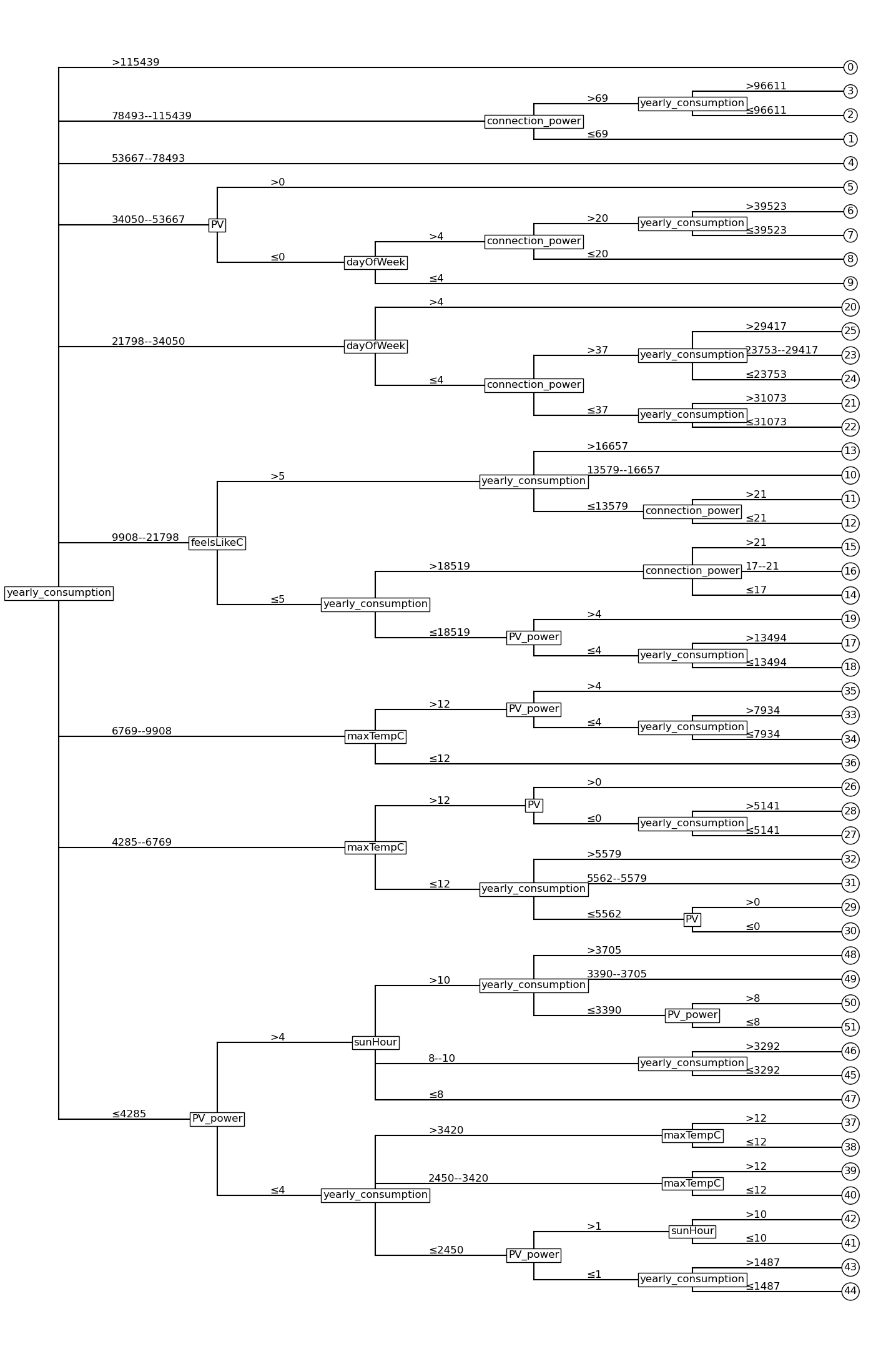}
    \caption{The tree learned with predictive clustering on the Flanders dataset}
    \label{fig:PCT_tree}
\end{figure*}

To illustrate the interpretability of the predictive clustering model, we show a PCT trained on the Flanders dataset\footnotemark in Fig.~\ref{fig:PCT_tree}.
\footnotetext{The trees for the other two datasets are available through \url{https://github.com/jonassoenen/predclus_scengen}. }
The tree itself is learned with a maximum depth of 7 to ensure that the full tree fits on a single page.
The tree is also compressed such that multiple consecutive binary splits on the same attribute become a single n-ary split. 
The compressed tree is equivalent to the original tree but is easier to interpret. 
Moreover, because each node in the tree corresponds to a cluster of time series, it is easy to see how a certain split influences the predictions of the model. 

From this tree (Fig.~\ref{fig:PCT_tree}), it is clear that yearly consumption is by far the most used attribute in the tree. 
This is unsurprising as yearly consumption provides direct information about the absolute consumption of the profile.
The other used attributes are the PV capacity, sun hour, temperature, day of the week, and connection power. 
Not all of these are used in the same subtrees: sun hour is only used to make splits for profiles that have PV panels reflecting that if you have PV panels, the sun hour attribute has a big influence on the consumption time series. 
On the other hand, connection power is a feature that is only used for consumers with high yearly consumption. 
Splitting on connection power might help to distinguish residential customers from small and medium enterprises.

\begin{figure*}
    \centering
    \includegraphics[width =.8\textwidth]{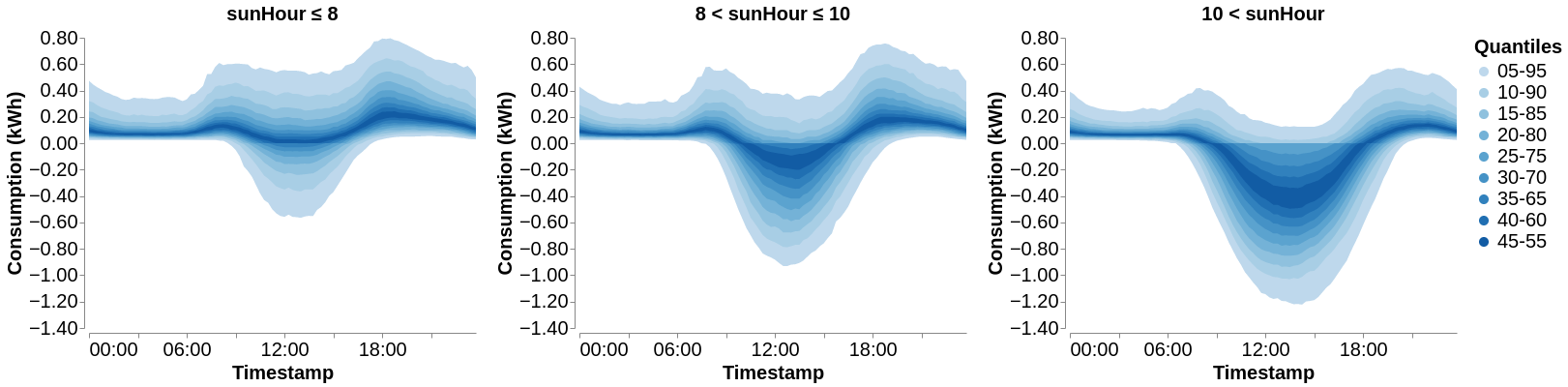}
    \caption{Visualization of the influence of the split on sunHour for consumers with a yearly consumption $\leq 4285$ kWh and PV power $> 4$ kVA for the Flanders dataset. Quantiles of all time series in the cluster are shown.}
    \label{fig:PCT_tree_vis_1}
\end{figure*}

\begin{figure}
    \centering
    \includegraphics[width =\columnwidth, trim={0 0 0 .5cm}]{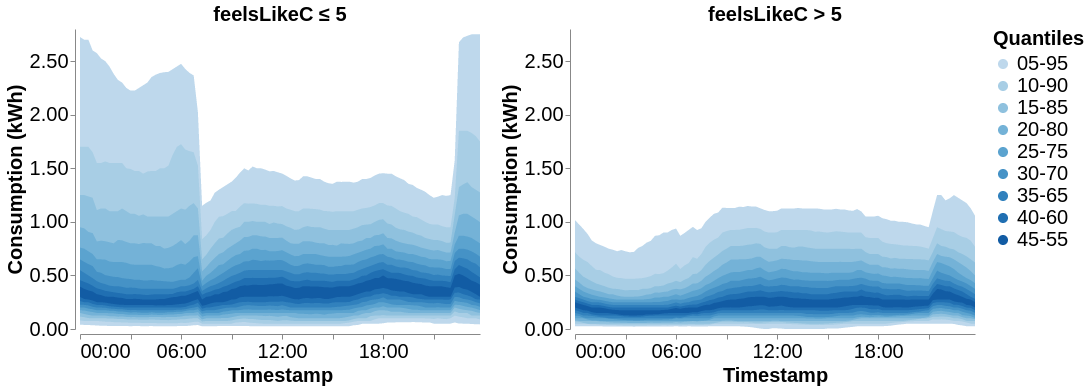}
    \caption{Visualization of the influence of the split on feelsLikeC for consumers with a yearly consumption in between 9908 kWh and 21798 kWh for the Flanders dataset. Quantiles of all time series in the cluster are shown.}
    \label{fig:PCT_tree_vis_2}
\end{figure}

Not only can this tree be inspected to see how the predictions are made, but by looking at all the time series in a leaf (or any internal node), one can also inspect what the model predicts in a certain branch of the tree or what influence a certain split has on the predictions. 
To visualize the time series in a node, we calculate the $0.05, 0.1, \dots, 0.90, 0.95$ quantiles and show those to give an idea of the distribution of the consumption values. 
In Fig.~\ref{fig:PCT_tree_vis_1}, we show the influence of the split on sun hour for consumers with a yearly consumption $\leq$ 4285 kWh that have a PV panel installation with a power $>$ 4 kVA.
In Fig.~\ref{fig:PCT_tree_vis_2}, we show the influence of the split on feelsLikeC temperature for consumers with a yearly consumption between 9907 and 21798 kWh. 
From the time series, it seems as if the consumers with a yearly consumption in this range are using accumulation heating with a tariff scheme that favors night consumption. 
When it is cold, accumulation heating draws power during the night in order to slowly release the created heat during the day. 
The PCT has learned that if feelsLikeC is lower than 5, significantly more electricity is needed for heating. The visualisations in this section were discussed with engineers and data scientists from Fluvius (the Flemish DSO), who agreed that they are insightful.

\section{Conclusion}
\label{sec:conclusion} 
In this paper, we have proposed a technique based on predictive clustering trees to generate electricity consumption scenarios for individual consumers based on their available attributes and attributes about the circumstances (such as weather and calendar information). 
Predictive clustering trees perform similarly to or slightly outperform the state-of-the-art data-driven method proposed in \cite{Soenen2023} while being significantly faster. 
Most importantly, the predictive clustering tree can be easily visualized such that a domain expert can observe which predictions are made in which circumstances, enabling the expert to gain new insights and trust the generated scenarios more easily. As future work, it would be interesting to compare the PCT method to generative neural network methods that are conditioned on the same variables.

\section{Acknowledgements} 
 This research received funding from the Flemish Government under the ``Onderzoeksprogramma Artifici\"ele Intelligentie (AI) Vlaanderen'' programme. 
The authors would like to thank the Flemish DSO Fluvius for supporting this work and discussing the interpretability of the method.
The load measurement dataset from the CER electricity consumption behaviour trial was accessed via the Irish Social Science Data Archive -\url{www.ucd.ie/issda}.

\bibliographystyle{IEEEtran}
\bibliography{references}

\begin{thebibliography}{10}
\providecommand{\url}[1]{#1}
\csname url@samestyle\endcsname
\providecommand{\newblock}{\relax}
\providecommand{\bibinfo}[2]{#2}
\providecommand{\BIBentrySTDinterwordspacing}{\spaceskip=0pt\relax}
\providecommand{\BIBentryALTinterwordstretchfactor}{4}
\providecommand{\BIBentryALTinterwordspacing}{\spaceskip=\fontdimen2\font plus
\BIBentryALTinterwordstretchfactor\fontdimen3\font minus \fontdimen4\font\relax}
\providecommand{\BIBforeignlanguage}[2]{{%
\expandafter\ifx\csname l@#1\endcsname\relax
\typeout{** WARNING: IEEEtran.bst: No hyphenation pattern has been}%
\typeout{** loaded for the language `#1'. Using the pattern for}%
\typeout{** the default language instead.}%
\else
\language=\csname l@#1\endcsname
\fi
#2}}
\providecommand{\BIBdecl}{\relax}
\BIBdecl

\bibitem{europeangreendeal}
``{EU responses to climate change},'' \url{https://www.europarl.europa.eu/news/en/headlines/society/20180703STO07129/eu-responses-to-climate-change}, 2022, [Online; accessed 14-Mar-2023].

\bibitem{europeangreendealenergy}
``{Energy and the Green Deal},'' \url{https://commission.europa.eu/strategy-and-policy/priorities-2019-2024/european-green-deal/energy-and-green-deal_en}, 2023, [Online; accessed 28-Mar-2023].

\bibitem{inflationreductionact}
``{Inflation Reduction Act Guidebook},'' \url{https://www.whitehouse.gov/cleanenergy/inflation-reduction-act-guidebook/}, 2022, [Online; accessed 14-Mar-2023].

\bibitem{inflationreductionactemissionreduction}
``{The Inflation Reduction Act Drives Significant Emissions Reductions and Positions America to Reach Our Climate Goals},'' \url{https://www.energy.gov/sites/default/files/202208/8.18\%20InflationReductionAct_Factsheet_Final.pdf}, 2022, [Online; accessed 14-Mar-2023].

\bibitem{neaimeh_deakin_jenkinson_giles_2023}
M.~Neaimeh, M.~Deakin, R.~Jenkinson, and O.~Giles, ``Democratizing electricity distribution network analysis,'' \emph{Data-Centric Engineering}, vol.~4, p.~e3, 2023.

\bibitem{ElementEnergy2022}
\BIBentryALTinterwordspacing
``Low voltage network capacity study,'' Element Energy and EA Technology, GOV.UK, Tech. Rep., 2022, accessed: 2023-04-19. [Online]. Available: \url{https://www.gov.uk/government/publications/low-voltage-network-capacity-study}
\BIBentrySTDinterwordspacing

\bibitem{Elmallah_2022}
S.~Elmallah, A.~M. Brockway, and D.~Callaway, ``Can distribution grid infrastructure accommodate residential electrification and electric vehicle adoption in northern california?'' \emph{Environmental Research: Infrastructure and Sustainability}, vol.~2, no.~4, p. 045005, nov 2022.

\bibitem{Vitiello2022a}
S.~Vitiello, N.~Andreadou, M.~Ardelean, and G.~Fulli, ``Smart metering roll-out in europe: Where do we stand? cost benefit analyses in the clean energy package and research trends in the green deal,'' \emph{Energies}, vol.~15, no.~7, 2022.

\bibitem{NavarroEspinosa2014}
A.~Navarro-Espinosa and P.~Mancarella, ``Probabilistic modeling and assessment of the impact of electric heat pumps on low voltage distribution networks,'' \emph{Applied Energy}, vol. 127, pp. 249--266, 2014.

\bibitem{NavarroEspinosa2016}
A.~Navarro-Espinosa and L.~F. Ochoa, ``Probabilistic impact assessment of low carbon technologies in lv distribution systems,'' \emph{IEEE Transactions on Power Systems}, vol.~31, no.~3, pp. 2192--2203, May 2016.

\bibitem{Veldman2015}
E.~Veldman and R.~A. Verzijlbergh, ``Distribution grid impacts of smart electric vehicle charging from different perspectives,'' \emph{IEEE Transactions on Smart Grid}, vol.~6, no.~1, pp. 333--342, Jan 2015.

\bibitem{Bernards2017}
R.~Bernards, J.~Morren, and H.~Slootweg, ``Statistical modelling of load profiles incorporating correlations using copula,'' in \emph{2017 IEEE PES Innovative Smart Grid Technologies Conference Europe (ISGT-Europe)}, Sep. 2017, pp. 1--6.

\bibitem{Bernards2020}
R.~Bernards, W.~van Westering, J.~Morren, and H.~Slootweg, ``Analysis of energy transition impact on the low-voltage network using stochastic load and generation models,'' \emph{Energies}, vol.~13, no.~22, 2020.

\bibitem{Soenen2023}
J.~Soenen, A.~Yurtman, T.~Becker, R.~D’hulst, K.~Vanthournout, W.~Meert, and H.~Blockeel, ``Scenario generation of residential electricity consumption through sampling of historical data,'' \emph{Sustainable Energy, Grids and Networks}, vol.~34, p. 100985, 2023.

\bibitem{Azam2023}
M.~F. Azam, T.~Becker, C.~Hermans, K.~Vanthournout, and G.~Deconinck, ``Optimized sampling strategy for load scenario generation in partially observable distribution grids,'' in \emph{2023 IEEE Belgrade PowerTech}, 2023, pp. 1--6.

\bibitem{Shi2018}
H.~Shi, M.~Xu, and R.~Li, ``Deep learning for household load forecasting—a novel pooling deep rnn,'' \emph{IEEE Transactions on Smart Grid}, vol.~9, no.~5, pp. 5271--5280, Sep. 2018.

\bibitem{Reis2017}
M.~Reis, A.~Garcia, and R.~J. Bessa, ``A scalable load forecasting system for low voltage grids,'' in \emph{2017 IEEE Manchester PowerTech}, June 2017, pp. 1--6.

\bibitem{Vossen2018}
J.~Vossen, B.~Feron, and A.~Monti, ``Probabilistic forecasting of household electrical load using artificial neural networks,'' in \emph{2018 IEEE International Conference on Probabilistic Methods Applied to Power Systems (PMAPS)}, June 2018, pp. 1--6.

\bibitem{Proedrou2021}
E.~Proedrou, ``A comprehensive review of residential electricity load profile models,'' \emph{IEEE Access}, vol.~9, pp. 12\,114--12\,133, 2021.

\bibitem{Swan2009}
L.~G. Swan and V.~I. Ugursal, ``Modeling of end-use energy consumption in the residential sector: A review of modeling techniques,'' \emph{Renewable and Sustainable Energy Reviews}, vol.~13, no.~8, pp. 1819--1835, 2009.

\bibitem{Grandjean2012}
A.~Grandjean, J.~Adnot, and G.~Binet, ``A review and an analysis of the residential electric load curve models,'' \emph{Renewable and Sustainable Energy Reviews}, vol.~16, no.~9, pp. 6539--6565, 2012.

\bibitem{Capasso1994}
A.~Capasso, W.~Grattieri, R.~Lamedica, and A.~Prudenzi, ``A bottom-up approach to residential load modeling,'' \emph{IEEE Transactions on Power Systems}, vol.~9, no.~2, pp. 957--964, May 1994.

\bibitem{Gottwalt2011}
S.~Gottwalt, W.~Ketter, C.~Block, J.~Collins, and C.~Weinhardt, ``Demand side management—a simulation of household behavior under variable prices,'' \emph{Energy Policy}, vol.~39, no.~12, pp. 8163--8174, 2011, clean Cooking Fuels and Technologies in Developing Economies.

\bibitem{McLoughlin2015}
F.~McLoughlin, A.~Duffy, and M.~Conlon, ``A clustering approach to domestic electricity load profile characterisation using smart metering data,'' \emph{Applied Energy}, vol. 141, pp. 190--199, 2015.

\bibitem{toubeau2016}
J.-F. Toubeau, M.~Hupez, V.~Klonari, Z.~De~Grève, and F.~Vallée, ``Statistical load and generation modelling for long term studies of low voltage networks in presence of sparse smart metering data,'' in \emph{IECON 2016 - 42nd Annual Conference of the IEEE Industrial Electronics Society}, Oct 2016, pp. 3900--3905.

\bibitem{Labeeuw2013}
W.~Labeeuw and G.~Deconinck, ``Residential electrical load model based on mixture model clustering and markov models,'' \emph{IEEE Transactions on Industrial Informatics}, vol.~9, no.~3, pp. 1561--1569, Aug 2013.

\bibitem{Johnson2014}
B.~J. Johnson, M.~R. Starke, O.~A. Abdelaziz, R.~K. Jackson, and L.~M. Tolbert, ``A matlab based occupant driven dynamic model for predicting residential power demand,'' in \emph{2014 IEEE PES T\&D Conference and Exposition}, 2014, pp. 1--5.

\bibitem{Bartels1992}
R.~Bartels, D.~G. Fiebig, M.~Garben, and R.~Lumsdaine, ``An end-use electricity load simulation model: Delmod,'' \emph{Utilities Policy}, vol.~2, no.~1, pp. 71--82, 1992.

\bibitem{LI2022124694}
\BIBentryALTinterwordspacing
J.~Li, Z.~Chen, L.~Cheng, and X.~Liu, ``Energy data generation with wasserstein deep convolutional generative adversarial networks,'' \emph{Energy}, vol. 257, p. 124694, 2022. [Online]. Available: \url{https://www.sciencedirect.com/science/article/pii/S0360544222015973}
\BIBentrySTDinterwordspacing

\bibitem{CHEN2023120711}
\BIBentryALTinterwordspacing
Z.~Chen, J.~Li, L.~Cheng, and X.~Liu, ``Federated-wdcgan: A federated smart meter data sharing framework for privacy preservation,'' \emph{Applied Energy}, vol. 334, p. 120711, 2023. [Online]. Available: \url{https://www.sciencedirect.com/science/article/pii/S0306261923000752}
\BIBentrySTDinterwordspacing

\bibitem{8791575}
Y.~Gu, Q.~Chen, K.~Liu, L.~Xie, and C.~Kang, ``Gan-based model for residential load generation considering typical consumption patterns,'' in \emph{2019 IEEE Power \& Energy Society Innovative Smart Grid Technologies Conference (ISGT)}, 2019, pp. 1--5.

\bibitem{WANG2020110299}
\BIBentryALTinterwordspacing
Z.~Wang and T.~Hong, ``Generating realistic building electrical load profiles through the generative adversarial network (gan),'' \emph{Energy and Buildings}, vol. 224, p. 110299, 2020. [Online]. Available: \url{https://www.sciencedirect.com/science/article/pii/S0378778820307234}
\BIBentrySTDinterwordspacing

\bibitem{CLAEYS2024122831}
\BIBentryALTinterwordspacing
R.~Claeys, R.~Cleenwerck, J.~Knockaert, and J.~Desmet, ``Capturing multiscale temporal dynamics in synthetic residential load profiles through generative adversarial networks (gans),'' \emph{Applied Energy}, vol. 360, p. 122831, 2024. [Online]. Available: \url{https://www.sciencedirect.com/science/article/pii/S0306261924002149}
\BIBentrySTDinterwordspacing

\bibitem{faraday}
S.~Chai and G.~Chadney, ``Faraday: Synthetic smart meter generator for the smart grid,'' in \emph{ICLR}.\hskip 1em plus 0.5em minus 0.4em\relax ICLR, 2024, pp. 1--6.

\bibitem{Blockeel1998}
H.~Blockeel, L.~D. Raedt, and J.~Ramon, ``Top-down induction of clustering trees,'' in \emph{Proceedings of the Fifteenth International Conference on Machine Learning}, ser. ICML '98.\hskip 1em plus 0.5em minus 0.4em\relax San Francisco, CA, USA: Morgan Kaufmann Publishers Inc., 1998, p. 55–63.

\bibitem{Dzeroski2007}
S.~D{\v{z}}eroski, V.~Gjorgjioski, I.~Slavkov, and J.~Struyf, ``Analysis of time series data with predictive clustering trees,'' in \emph{Knowledge Discovery in Inductive Databases}, S.~D{\v{z}}eroski and J.~Struyf, Eds.\hskip 1em plus 0.5em minus 0.4em\relax Berlin, Heidelberg: Springer Berlin Heidelberg, 2007, pp. 63--80.

\bibitem{quinlan2014}
J.~R. Quinlan, \emph{C4. 5: programs for machine learning}.\hskip 1em plus 0.5em minus 0.4em\relax Elsevier, 2014.

\bibitem{Sakoe1978}
H.~Sakoe and S.~Chiba, ``Dynamic programming algorithm optimization for spoken word recognition,'' \emph{IEEE Transactions on Acoustics, Speech, and Signal Processing}, vol.~26, no.~1, pp. 43--49, 1978.

\bibitem{Petitjean2011a}
F.~Petitjean, A.~Ketterlin, and P.~Gançarski, ``A global averaging method for dynamic time warping, with applications to clustering,'' \emph{Pattern Recognition}, vol.~44, no.~3, pp. 678--693, 2011.

\bibitem{Quinlan1987}
J.~R. Quinlan, ``Simplifying decision trees,'' \emph{Int. J. Man Mach. Stud.}, vol.~27, pp. 221--234, 1987.

\bibitem{irishdata}
{Commission for Energy Regulation (CER)}, ``{CER Smart Metering Project - Electricity Consumer Behaviour Trial, 2009-2010 },'' Irish Social Science Data Archive, [dataset] 1st Edition, SN: 0012-00, 2012, \url{https://www.ucd.ie/issda/data/commissionforenergyregulationcer/}.

\bibitem{londondata}
{Tindemans, S., Strbac, G., Schofield, J. R, Woolf, M., Carmichael, R., Bilton, M.}, ``Low carbon london project: Data from the dynamic time-of-use electricity pricing trial, 2013,'' UK Data Service, [data collection] SN: 7857, 2016, \url{http://doi.org/10.5255/UKDA-SN-7857-2}.

\bibitem{leakage}
S.~Kaufman, S.~Rosset, and C.~Perlich, ``Leakage in data mining: Formulation, detection, and avoidance,'' in \emph{Proceedings of the 17th ACM SIGKDD International Conference on Knowledge Discovery and Data Mining}, ser. KDD '11.\hskip 1em plus 0.5em minus 0.4em\relax New York, NY, USA: Association for Computing Machinery, 2011, p. 556–563.

\bibitem{Haben2014}
S.~Haben, J.~Ward, D.~Vukadinovic~Greetham, C.~Singleton, and P.~Grindrod, ``{A new error measure for forecasts of household-level, high resolution electrical energy consumption},'' \emph{International Journal of Forecasting}, vol.~30, no.~2, pp. 246--256, 2014.

\bibitem{Gneiting2008}
T.~Gneiting, L.~Stanberry, E.~Grimit, L.~Held, and N.~Johnson, ``Assessing probabilistic forecasts of multivariate quantities, with an application to ensemble predictions of surface winds,'' \emph{TEST}, vol.~17, no.~2, pp. 211--235, 7 2008.

\bibitem{Moehrlen2023}
C.~Möhrlen, J.~W. Zack, and G.~Giebel, ``Chapter fourteen - assessment of forecast performance,'' in \emph{IEA Wind Recommended Practice for the Implementation of Renewable Energy Forecasting Solutions}, ser. Wind Energy Engineering, C.~Möhrlen, J.~W. Zack, and G.~Giebel, Eds.\hskip 1em plus 0.5em minus 0.4em\relax Academic Press, 2023, pp. 125--145.

\bibitem{Matheson1976}
J.~Matheson and R.~Winkler, ``Scoring rules for continuous probability distributions,'' \emph{Management Science}, vol.~22, no.~10, pp. 1087--1096, 6 1976.

\bibitem{Broecker2007}
J.~Bröcker and L.~A. Smith, ``Scoring probabilistic forecasts: The importance of being proper,'' \emph{Weather and Forecasting}, vol.~22, no.~2, pp. 382 -- 388, 2007.

\bibitem{Satopaa2011}
V.~Satopaa, J.~Albrecht, D.~Irwin, and B.~Raghavan, ``Finding a "kneedle" in a haystack: Detecting knee points in system behavior,'' in \emph{2011 31st International Conference on Distributed Computing Systems Workshops}, June 2011, pp. 166--171.

\bibitem{Breiman2001}
L.~Breiman, ``Random forests,'' \emph{Machine learning}, vol.~45, no.~1, pp. 5--32, 2001.

\bibitem{Pedregosa2011}
F.~Pedregosa, G.~Varoquaux, A.~Gramfort, V.~Michel, B.~Thirion, O.~Grisel, M.~Blondel, P.~Prettenhofer, R.~Weiss, V.~Dubourg, J.~Vanderplas, A.~Passos, D.~Cournapeau, M.~Brucher, M.~Perrot, and E.~Duchesnay, ``Scikit-learn: Machine learning in {P}ython,'' \emph{Journal of Machine Learning Research}, vol.~12, pp. 2825--2830, 2011.

\end{thebibliography}

\end{document}